\documentclass[aps,prl,twocolumn,superscriptaddress,longbibliography]{revtex4-1}

\usepackage{hyperref} 
\usepackage{graphicx}
\usepackage{bm}
\usepackage{amsmath}
\usepackage{amssymb}

\DeclareGraphicsExtensions{.png,.jpg,.eps}

\begin{document}
\author{Wei Luo}
\affiliation{National Laboratory of Solid State Microstructures and School of Physics, Nanjing University, Nanjing 210093, China}
\affiliation{School of Science, Jiangxi University of Science and Technology, Ganzhou 341000, China}

\author{Wei Chen}
\email{Corresponding author: pchenweis@gmail.com}
\affiliation{National Laboratory of Solid State Microstructures and School of Physics, Nanjing University, Nanjing 210093, China}
\affiliation{Collaborative Innovation Center of Advanced Microstructures, Nanjing University, Nanjing 210093, China}

\author{D. Y. Xing}
\affiliation{National Laboratory of Solid State Microstructures and School of Physics, Nanjing University, Nanjing 210093, China}
\affiliation{Collaborative Innovation Center of Advanced Microstructures, Nanjing University, Nanjing 210093, China}

\title{Anomalous Andreev Reflection on a Torus-Shaped Fermi Surface}

\begin{abstract}
Andreev reflection (AR) refers to the electron-hole conversion at the normal metal-superconductor interface.
In a three-dimensional metal with spherical Fermi surface, retro (specular) AR can occur with the sign reversal
of all three (a single) components of particle velocity. Here, we predict a novel type of AR
with the inversion of two velocity components, dubbed ``anomalous-trajectory Andreev reflection'' (AAR), which can be realized
in a class of materials with  torus-shaped Fermi surface,
such as doped nodal line semimetals. For its toroidal circle perpendicular to the interface,
the Fermi torus doubles the AR channels and generates multiple AR processes.
In particular, the AAR and retro AR are found to dominate electron transport
in the light and heavy doping regimes, respectively.
We show that the AAR visibly manifests as a ridge structure
in the spatially resolved nonlocal conductance,
in contrast to the peak structure for the retro AR.
Our work opens a new avenue for the AR spectroscopy
and offers a clear transport signature of torus-shaped Fermi surface.
\end{abstract}

\date{\today}

\maketitle

Andreev reflection (AR) describes the electron-hole
conversion at the interface between a normal metal and a
superconductor, by which
a pair of electrons penetrate into the superconductor
and form a Cooper pair \cite{Andreev64jetp}. It dominates the electron
transport within the superconducting gap
due to the prohibition of single-particle
transmission \cite{Blonder82prb}.
AR spectroscopy has become a powerful
tool for the detection
of various properties of electronic systems, such as
the pairing order parameter in unconventional superconductors \cite{Bruder90prb,Tanaka95prl,Kashiwaya00rpp},
spin polarization of metal \cite{Jong95prl,Soulen98sci},
Dirac/Weyl fermion with linear dispersion \cite{Beenakker06prl,Zhang08prl,Chen13epl},
and Majorana zero mode in topological superconductors
\cite{Law09prl,Mourik12sci}.
It may also have potential
applications in quantum information
processing \cite{Recher01prb,Lesovik01epjb,Hofstetter09nat} and spintronics \cite{Eschrig11pt}.

\begin{figure}[t!]
 \centering
   \includegraphics[width=1\columnwidth]{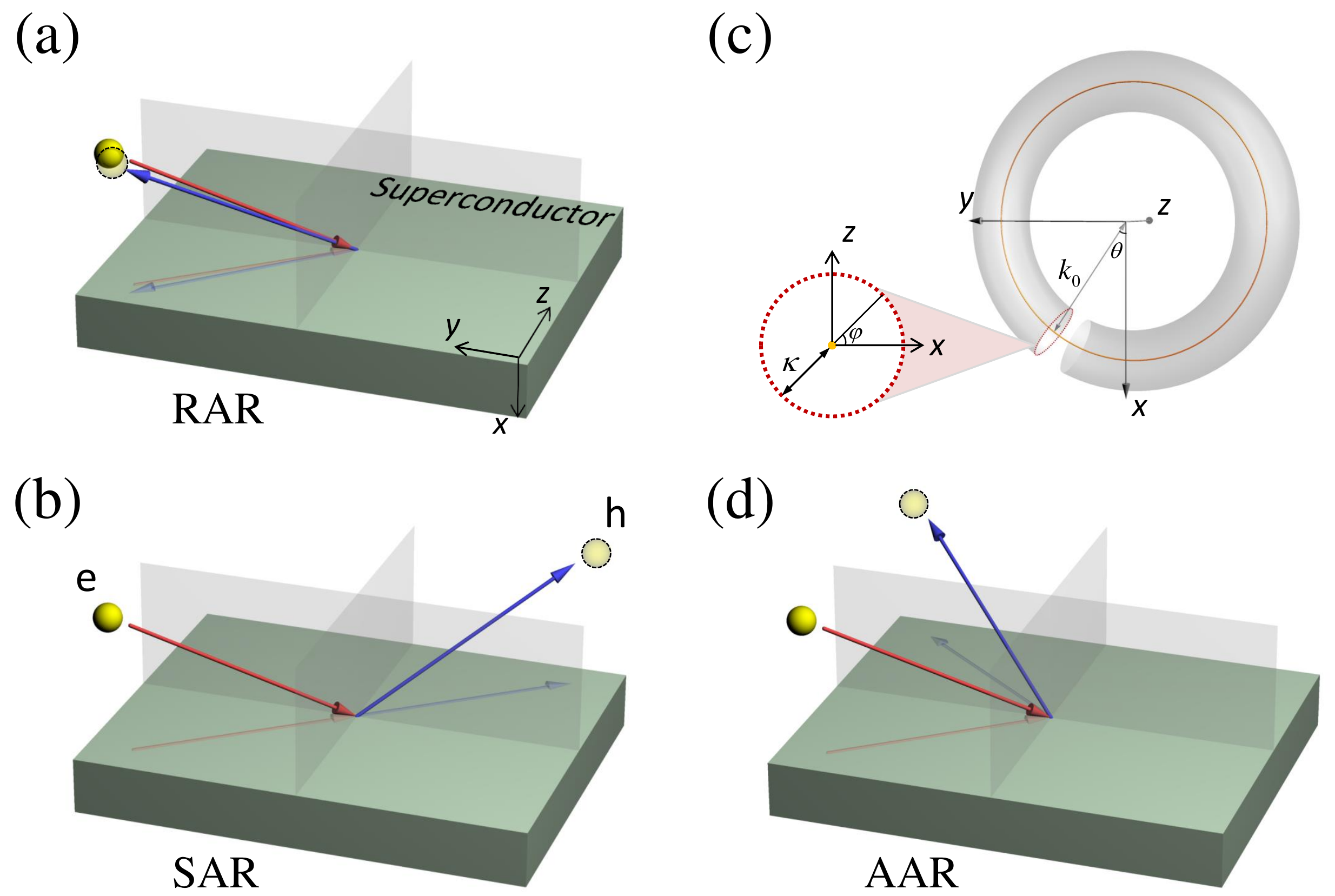}
\caption{Schematic illustration of three types of AR, an
electron (e) incident from the normal metal is reflected into a hole (h):
(a) RAR, (b) SAR, and
(d) AAR with three, one and two
velocity components reversed relative
to the incident electron. The projected trajectories
on the $y$-$z$ plane are indicated.
(c) Torus-shaped Fermi surface of a doped nodal line
semimetal with major radius $k_0$, minor redius $\kappa$, toroidal angle $\theta$, and poloidal angle $\varphi$.}
\label{Fig1}
\end{figure}

One unique feature of the conventional AR
is that the reflected hole retraces
the path of the incident electron by
inverting all velocity components \cite{Andreev64jetp},
so-called retro AR (RAR); see Fig. \ref{Fig1}(a).
Interestingly, if the incident electron and reflected
hole are from the conduction and valence band respectively,
specular AR (SAR) may occur \cite{Beenakker06prl},
in which only the
velocity component perpendicular to the interface
is inverted; see Fig. \ref{Fig1}(b). The SAR
is hard to achieve in conventional
semiconductors, whose energy gap exceeds the
superconducting gap so as to
block the reflected hole.
However, massless Dirac Fermion
in various materials, such as graphene and
topological semimetals is favorable for the observation of SAR \cite{Beenakker06prl,Chen13epl,Efetov16np}.
In two dimensions, the RAR and SAR constitute a
complete set of AR. Interestingly,
the phenomenon gets enriched in three dimensions
due to the additional velocity component.

\begin{table}[htb]
\centering
\caption{Comparison of
velocity inversion
in the RAR, SAR, and AAR.}\label{Tab:Comparison}
\begin{ruledtabular}
\begin{tabular}{llll}
& $v_x$  & $v_y$ & $v_z$ \\ \hline
RAR & $-$  & $-$ & $-$ \\
SAR & $-$ & + & + \\
AAR & $-$ & $-$ (+) & + ($-$) \\
\end{tabular}
\end{ruledtabular}
\end{table}

In this Letter, we predict the third
type of AR in which
two of three velocity components are inverted,
dubbed ``anomalous-trajectory
Andreev reflection'' (AAR); see Fig. \ref{Fig1}(d)
and Table \ref{Tab:Comparison} for clarity.
We show that the AAR can occur in the system
with a torus-shaped Fermi surface
\cite{Emmanouilidou17prb,Takane18npjqm,Hirose20prb,Kwan20prr} as shown in Fig. \ref{Fig1}(c),
which exists in a large category of materials
called nodal line semimetals \cite{Burkov11prb,Kim15prl,Yu15prl,
Heikkila11jetp,Weng15prb,Chen15nl,Zeng15arxiv,Fang15prb,Yamakage16jpsj,
Xie15aplm,Chan16prb,Zhao16prb,Bian16prb,Bian16nc,Emmanouilidou17prb,
Takane18npjqm,Hirose20prb,Kwan20prr}.
The Fermi torus differs in topology from the conventional Fermi
sphere, which can lead to anomalous transport signatures \cite{Chen19prl,Li18prl}.
Here, we consider a metal-superconductor junction oriented along the
$x$-direction and the toroidal circle of the Fermi torus
is perpendicular to the interface. For lateral momentum $(k_y, k_z)$ conserved,
there exist two AR channels
with different sign in the lateral velocities
[cf. Table \ref{Tab:Comparison}].
Specifically, we find that the AAR dominates
the inter-band scattering regime in which the electron and reflected hole
are from different bands,
as the sample is weakly doped such that
the chemical potential is much smaller than the superconducting gap
($\mu\ll\Delta$).
The main feature of AAR is that only one
lateral velocity component inverts its
sign, indicating an unconventional
trajectory of the reflected hole wave packet,
that is, extended in one direction while
localized in the other.
Such a feature can be probed by nonlocal scanning tunneling
spectroscopy (STM), in which the conductance
exhibits a spatially resolved ridge structure,
thus providing a clear signature of AAR.

We consider a spin-degenerate nodal line semimetal
captured by the following Hamiltonian
\begin{equation}\label{NL}
H_{\text{NL}}=\hbar\lambda(k_x^2+k_y^2-k_0^2)\sigma_x+\hbar v k_z\sigma_y-\mu,
\end{equation}
where the Pauli matrices $\sigma_{x,y}$ operate on the pseudo-spin (orbital/sublattice),
$k_0$ is the radius of the nodal loop defined by the band crossing,
$\lambda$ and $v$ are model parameters, and $\mu$
is the chemical potential.
In order to study the electron-hole scattering,
we work on the Bogoliubov-de Gennes Hamiltonian
for the whole junction
\begin{equation}\label{bdg}
\begin{split}
\mathcal{H}&=\mathcal{H}_{\text{NL}}\Theta(-x)+\mathcal{H}_{\text{S}}\Theta(x)+U\delta(x)\tau_z,\\
\mathcal{H}_{\text{NL}}&=H_{\text{NL}}\otimes\tau_z, \ \ \
\mathcal{H}_{\text{S}}=\varepsilon_{\bm{k}}\sigma_0\otimes\tau_z+\Delta\sigma_0\otimes\tau_x,
\end{split}
\end{equation}
where $\Theta(x)$ is the unit step function and $\tau_{x,z}$
are Pauli matrices in the Nambu space,
$\mathcal{H}_{\text{NL}}$ and $\mathcal{H}_{\text{S}}$
correspond to the nodal line semimetal and the conventional $s$-wave superconductor, respectively,
and $U\delta(x)\tau_z$ is a $\delta$-function interface barrier at $x=0$.
The normal state energy $\varepsilon_{\bm{k}}=\hbar^2\bm{k}^2/(2m)-\mu_s$
(relative to the chemical potential $\mu_s$) in the superconductor
and the pair potential $\Delta$ are
diagonal in the pseudo-spin space, and $m$
is the effective mass.

\begin{figure}[t!]
 \centering
   \includegraphics[width=1\columnwidth]{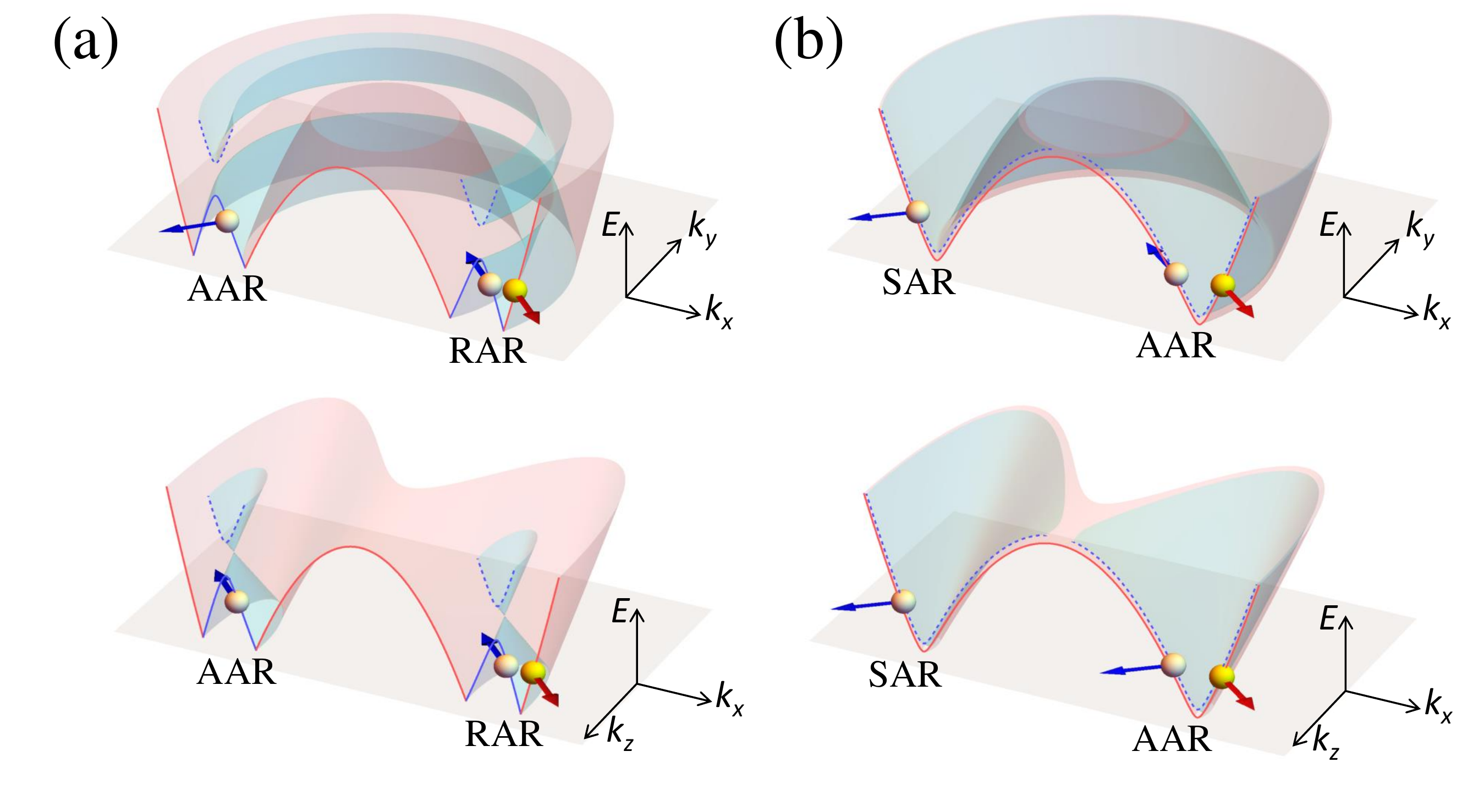}
\caption{
The dispersion of electron (red band) and hole (blue band)
with (a) finite and (b) vanishing chemical potential,
in which solid and dashed boundaries denote
conduction and valence bands, respectively. For an incident
electron (red arrow), there are (a) two intra-band AR channels
for the hole (blue arrows): RAR and AAR,
and (b) two inter-band AR channels: AAR and SAR.
}
\label{Fig2}
\end{figure}

For a slightly doped nodal line semimetal, in which
$\mu\ll\hbar\lambda k_0^2$ and
the Fermi surface possesses a torus shape,
we can linearize and parameterize
the Hamiltonian as $\mathcal{H}_{\text{NL}}\simeq[\hbar v_0\kappa
(\cos\varphi\sigma_x+\sin\varphi\sigma_y)-\mu]\otimes\tau_z$
around the nodal loop
by substituting $k_x=(k_0+\kappa\cos\varphi)\cos\theta$,
$k_y=(k_0+\kappa\cos\varphi)\sin\theta$ and $k_z=\kappa\sin\varphi/\alpha$,
where $\alpha=v/v_0$ is the ratio between the velocities in the $z$ direction
and $v_0=2\lambda k_0$ in the $x$-$y$ plane.
All states are then specified by the
minor radius $\kappa$, toroidal angle $\theta$, and poloidal
angle $\varphi$ of the torus; see Fig. \ref{Fig1}(c).
The excitation energies for electron and hole
are $E_e=\pm\hbar v_0\kappa-\mu$
and $E_h=\mp\hbar v_0\kappa+\mu$, which are
independent of $\theta$ and $\varphi$, with ``$\pm$'' (``$\mp$'')
corresponding to the
conduction and valence band, respectively.

The Fermi torus has
nontrivial effects on AR by
doubling the reflection channels.
Consider an electron in the conduction band incident on the interface at $x= 0$ from the normal side,
with energy $E$ and transverse wave vector $(k_y, k_z)$, or
equivalently labeled by
$(\kappa_e, \theta_e, \varphi_e)$, with $\kappa_e=(E+\mu)/(\hbar v_0)$.
There are two possible incident states ($k_x>0$ or $k_x<0$)
on the Fermi torus, and we first focus on the case of
$k_x>0$ or $\theta_e\in(-\pi/2,\pi/2)$.
The velocity components are given by
$v^e_x=v_0\cos\varphi_e\cos\theta_e, v^e_y=v_0\cos\varphi_e\sin\theta_e$ and
$v^e_z=v\sin\varphi_e$.
There are two scenarios of AR
depending on the relative magnitude of $E$ and $\mu$.
(i) For $E<\mu$, the Andreev reflected hole is in
the conduction band, the same as the incident electron,
indicating an intra-band process; see Fig. \ref{Fig2}(a).
There are two reflecting hole states:
$h_1:(\kappa_h, \theta_h, \varphi_h)$ and $h_2
:(\kappa_h, \pi-\theta_h, \pi-\varphi_h)$
where $\kappa_h=|E-\mu|/(\hbar v_0)$ is the minor radius
of the hole torus, $\theta_h\simeq\theta_e$ up to a small correction
of the order $\kappa_{e(h)}/k_0$, and $\sin\varphi_h/\sin\varphi_e
=\kappa_e/\kappa_h$.
Since the dispersion of the hole in the
conduction band is opposite in sign to that for the electron,
the $h_1$ state inverts all three velocity
components, that is RAR.
For the $h_2$ state, different reflection angles
contribute additional minus sign to $v_y^h$,
so that only $v_x^h$ and $v_z^h$ are inverted in the end,
which is AAR. (ii) For $E>\mu$, the Andreev
reflected hole is in the valence band, which is
an inter-band process. A typical case of $\mu\simeq0$ is illustrated
in Fig. \ref{Fig2}(b).
Now the hole possesses the same sign in dispersion as the electron,
and two AR states
are $\tilde{h}_1:(\kappa_h, \pi-\theta_h, \varphi_h)$ and
$\tilde{h}_2: (\kappa_h, \theta_h, \pi-\varphi_h)$ by requiring
$v^h_x<0$. Correspondingly,
the hole state $\tilde{h}_1$ contains only
the sign reversal in $v_x^h$,
being the SAR. Meanwhile,
for the state $\tilde{h}_2$ both $v_x^h$ and $v_y^h$
change the sign [cf. Fig. \ref{Fig1}(d)], which is again AAR.
From the qualitative analysis above, one finds that all types of
AR
can occur on the Fermi torus, as summarized
in Table \ref{Tab:Comparison}.

Next we solve the scattering problem
in detail to give quantitative results. For an incident electron $e_1$ with $k_x>0$,
the wave functions on both sides of the junction are given by
\begin{equation}\label{wave}
\begin{split}
\psi&=\big[\psi_{\text{NL}}\Theta(-x)+\psi_{\text{S}}\Theta(x)\big]e^{ik_yy+ik_zz},\\
\psi_{\text{NL}}&=(e^{ik_1^ex}+r_{e_{1}}^{e_{1}}e^{-ik_1^ex})|e_1\rangle+r_{e_{1}}^{e_{2}}e^{ik^e_2x}|e_2\rangle\\
&\ \ \ +r_{e_{1}}^{h_{1}}e^{i\eta k_1^hx}|h_1\rangle+r_{e_{1}}^{h_{2}}e^{-i\eta k_2^hx}|h_2\rangle,\\
\psi_{\text{S}}&=(t_1|s_1\rangle+t_2|s_2\rangle)e^{iq_1x}+(t_3|s_3\rangle+t_4|s_4)\rangle e^{-iq_2x}.
\end{split}
\end{equation}
Here,  $\psi_{\text{NL}}$ denotes the wave function of the nodal line semimetal for $x<0$
and its spinor part is given by
$|e_1\rangle=(1/\sqrt{2},e^{i\varphi_e}/\sqrt{2},0,0)$,
$|e_2\rangle=(1/\sqrt{2},-e^{-i\varphi_e}/\sqrt{2},0,0)$,
$|h_1\rangle=(0,0,1/\sqrt{2},\eta e^{i\varphi_h}/\sqrt{2})$, and
$|h_2\rangle=(0,0,1/\sqrt{2},-\eta e^{-i\varphi_h}/\sqrt{2})$
with $\varphi_{e(h)}=\sin^{-1}[\alpha k_z/\kappa_{e(h)}]$
and $\eta=\text{sgn}(\mu-E)$. The $x$ components of the
wave vectors are $k_{1}^{e(h)}=[k_0+\kappa_{e(h)}\cos\varphi_{e(h)}]\cos\theta_{e(h)}^+$ and
$k_{2}^{e(h)}=[k_0-\kappa_{e(h)}\cos\varphi_{e(h)}]\cos\theta_{e(h)}^-$
with $\theta^\pm_{e(h)}=\sin^{-1}[k_y/(k_0\pm\kappa_{e(h)}\cos\varphi_{e(h)})]$.
The amplitudes $r_{e_{1}}^{e_{1}}$ and $r_{e_{1}}^{e_{2}}$ correspond to the normal reflection,
and $r_{e_{1}}^{h_{1}}$ and $r_{e_{1}}^{h_{2}}$ are marked as the RAR (SAR) and
AAR, respectively.
$\psi_{\text{S}}$ stands for the wave function of the superconductor for $x>0$, in which
$t_1,...,t_4$ are the quasiparticle tunneling amplitudes, and the wave vectors are
$q_{1,2}=[2m(\mu_s\pm\sqrt{E^2-\Delta^2})/\hbar^2-k_y^2-k_z^2]^{\frac{1}{2}}$.
The spinor part of $\psi_{\text{S}}$ is given by
$|s_1\rangle=|s_3\rangle^*=(1/\sqrt{2},0,e^{-i\beta}/\sqrt{2},0)$ and
$|{s}_2\rangle=|{s}_4\rangle^*=(0,1/\sqrt{2},0,e^{-i\beta}/\sqrt{2})$
with $\beta=\cos^{-1}(E/\Delta)$ for $E<\Delta$
and $i\cosh^{-1}{(E/\Delta)}$ for $E>\Delta$.
All scattering coefficients can be solved by boundary conditions,
$\psi_{\text{NL}}(0)=\psi_{\text{S}}(0)=\psi(0)$ and
$
\lambda \sigma_{x} \psi^{\prime}_{\text{NL}}(0)-\frac{\hbar^2}{2m}\sigma_{0} \psi_{\text{S}}^{\prime}(0)=-\sigma_{0}  U \psi(0).
$

\begin{figure}[t!]
 \centering
   \includegraphics[width=\columnwidth]{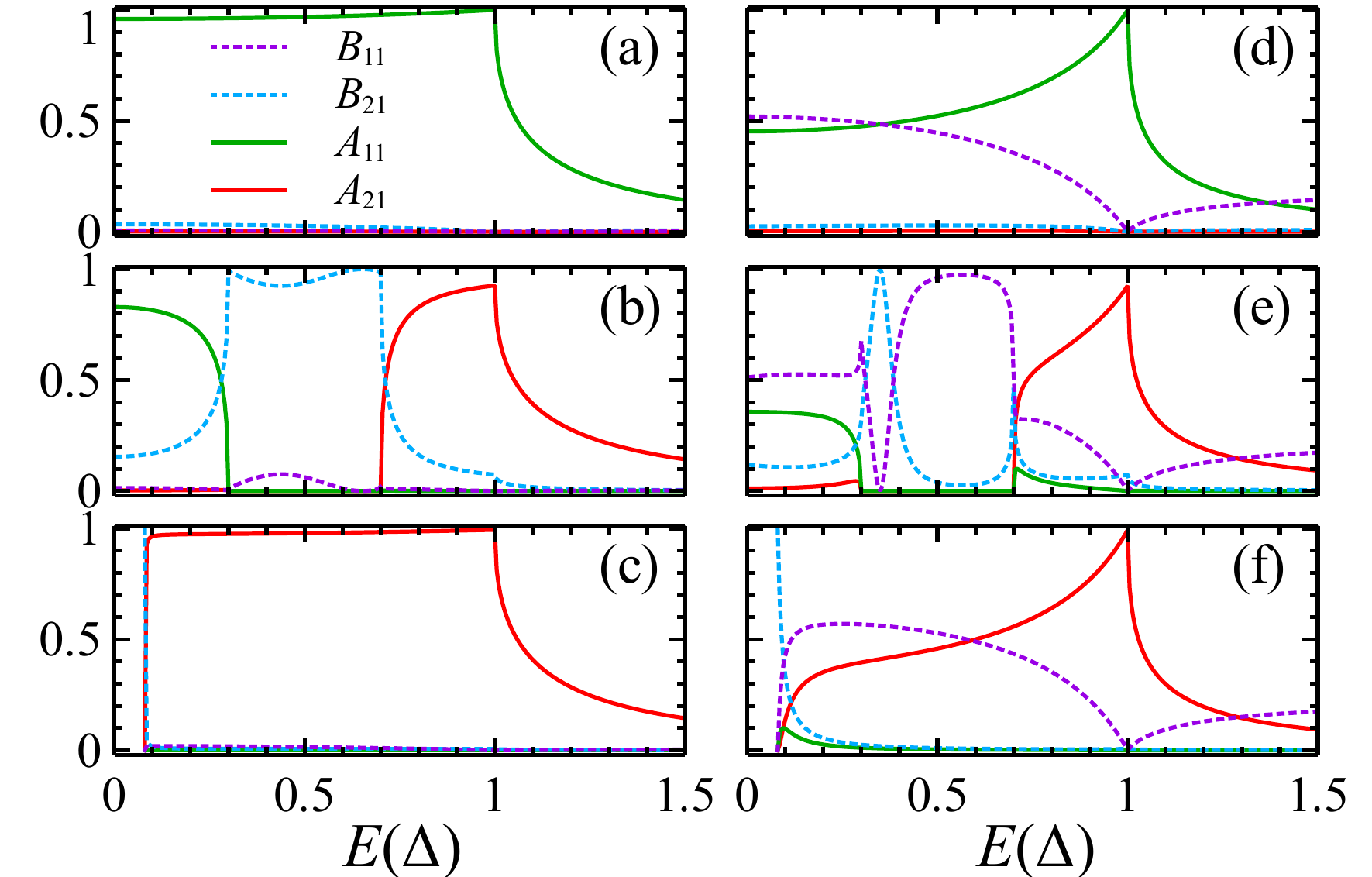}
\caption{Probabilities of various AR processes as
functions of energy. Specifically, $A_{21}$ is the AAR probability and $A_{11}$ corresponds to
RAR (SAR) in the intra (inter)-band regime. The interface barrier $Z=2U/(\hbar v_0) =0$  in
the left panel and $Z=1$ in the right panel.
The chemical potential and $k_z$ are taken as (a,d) $\mu=40\Delta$ and $ k_z=0.2k_0$,
(b,e) $\mu=0.5\Delta$ and $ k_z=k_0/200$, and (c,f) $\mu=0$ and $k_z=k_0/500$,
and $k_y=0.5k_0$ is adopted in (a-f).
The parameters for the superconductor are $\Delta=2$ meV, $\mu_{s}=2$ eV, $m=0.1m_e$
and those for the nodal line semimetal are $k_{0}=2$ nm$^{-1}$, $v_0=3\times10^5$ m/s,
$\alpha=0.2$.}
\label{Fig3}
\end{figure}

To see some analytical results, we first consider a transparent interface ($U=0$)
and two limiting cases: $E, \Delta\ll\mu$ and $E, \Delta\gg\mu$,
such that we have $\theta_{e}^{\pm}\simeq\theta_h^\pm=\theta$ and $\varphi_e\simeq\varphi_h=\varphi$.
More general numerical results will be given later.
For $E, \Delta\ll\mu$, the amplitudes $r_{e_{1}}^{h_{1}}$ for the RAR process and $r_{e_{1}}^{h_{2}}$ for the AAR process with
inverted $v_x$ and $v_z$ are obtained as
\begin{equation}\label{intra}
\begin{split}
r_{e_{1}}^{h_{1}}&=\frac{2\chi[2\chi\cos{\beta}\cos^{2}{\varphi}+i\sin{\beta}(\cos^{2}{\varphi}+\chi^{2})]}{[i\sin{\beta}(\cos^{2}\varphi+\chi^{2})+2\chi\cos{\beta}]^{2}-4\chi^{2}\sin^{2}{\varphi}},\\
r_{e_{1}}^{h_{2}}&=\frac{-2\chi e^{i\varphi}\sin{\beta}\sin{\varphi}(\cos^{2}{\varphi}-\chi^{2})}{[i\sin{\beta}(\cos^{2}\varphi+\chi^{2})+2\chi\cos{\beta}]^{2}-4\chi^{2}\sin^{2}{\varphi}},
\end{split}
\end{equation}
where
$\chi=v^e_x/v^{\text{F}}_x$ with $v^e_x$ as the $x$-direction velocity of electron in the nodal line semimetal
and $v_x^{\text{F}}=\sqrt{2m\mu_s-\hbar^2(k_y^2+k_z^2)}/m$ as that in the superconductor. In this case,
the RAR process dominates the electron transport for $E< \Delta$. For $E=\Delta$ and thus $\beta=0$, we have $r_{e_{1}}^{h_{1}}=1$ and $r_{e_{1}}^{h_{2}}=0$,
indicating that perfect RAR
can be implemented for all incident angles $(\theta, \varphi)$.

For $E, \Delta\gg\mu$,
the AR amplitudes are reduced to
\begin{equation}\label{inter}
\begin{split}
r_{e_{1}}^{h_{1}}=0,\ \ \
r_{e_{1}}^{h_{2}}=\frac{2\chi e^{i\varphi}\cos{\varphi}}
{i\sin{\beta}(\cos^{2}{\varphi}+\chi^{2})+2\chi\cos{\beta}},
\end{split}
\end{equation}
where
$r_{e_{1}}^{h_{1}}$ and $r_{e_{1}}^{h_{2}}$ correspond to the SAR and AAR processes, respectively. In this case,
one finds that the AAR dominates the electron transport with the inversion of $v_x$ and $v_y$. In particular,
the AAR amplitude becomes
$r_{e_{1}}^{h_{2}}=e^{i\varphi}\cos\varphi$ at
$E=\Delta$, solely determined by the
poloidal angle $\varphi$.
The vanishing SAR in Eq. \eqref{inter}
stems from the orthogonality between the pseudo-spin texture of the
incident and reflected waves, i.e., $\langle{h}_1|\tau_x|{e}_1\rangle=0$
for $\eta=-1$.  However, the existence of an interface
barrier can induce multiple scattering and lead to a finite SAR probability [cf. Fig. \ref{Fig3}(f)].
For the incident electron $e_2$ with $k_x<0$, the similar calculations for wave functions can be done, and
the corresponding AR amplitudes of $e_1$ ($k_x>0$) and $e_2$ ($k_x<0$) are shown to be related  by
$r_{e_{2}}^{h_{1}}(\varphi)=r_{e_{1}}^{h_{2}}(-\varphi)$
and
$r_{e_{2}}^{h_{2}}(\varphi)=r_{e_{1}}^{h_{1}}(-\varphi)$,
where $r_{e_{2}}^{h_{1}}$ is the AAR amplitude of the $e_2$ electron,
and $r_{e_{2}}^{h_{2}}$ corresponds to its RAR or SAR amplitude.

For more general cases, we plot the numerical results
of various reflection probabilities in
Fig. \ref{Fig3}, which are defined as
$B_{11}=|r_{e_{1}}^{e_{1}}|^2$, $B_{21}=|r_{e_{1}}^{e_{2}}|^2v^e_{2x}/v_{1x}^e$,
$A_{11}=|r_{e_{1}}^{h_{1}}|^2v^h_{1x}/v_{1x}^e$, and $A_{21}=|r_{e_{1}}^{h_{2}}|^2v^h_{2x}/v_{1x}^e$
with the ratio between the velocities of reflected and incident waves involved.
In Figs. \ref{Fig3}(a) and (c), the calculated
results recover the conclusions drawn from Eqs. \eqref{intra} and \eqref{inter}
in two limiting cases. Between the two, as shown in Fig. \ref{Fig3}(b), there is a crossover
from the intra-band RAR-dominated process to inter-band AAR-dominated one \cite{sm}.
 The effect of the interface barrier is considered in
Figs. \ref{Fig3}(d)-(f), which enhances the normal
electron reflection, but does not much change the
relative strength of various AR processes.

\begin{figure}[t!]
\centering
\includegraphics[width=1\columnwidth]{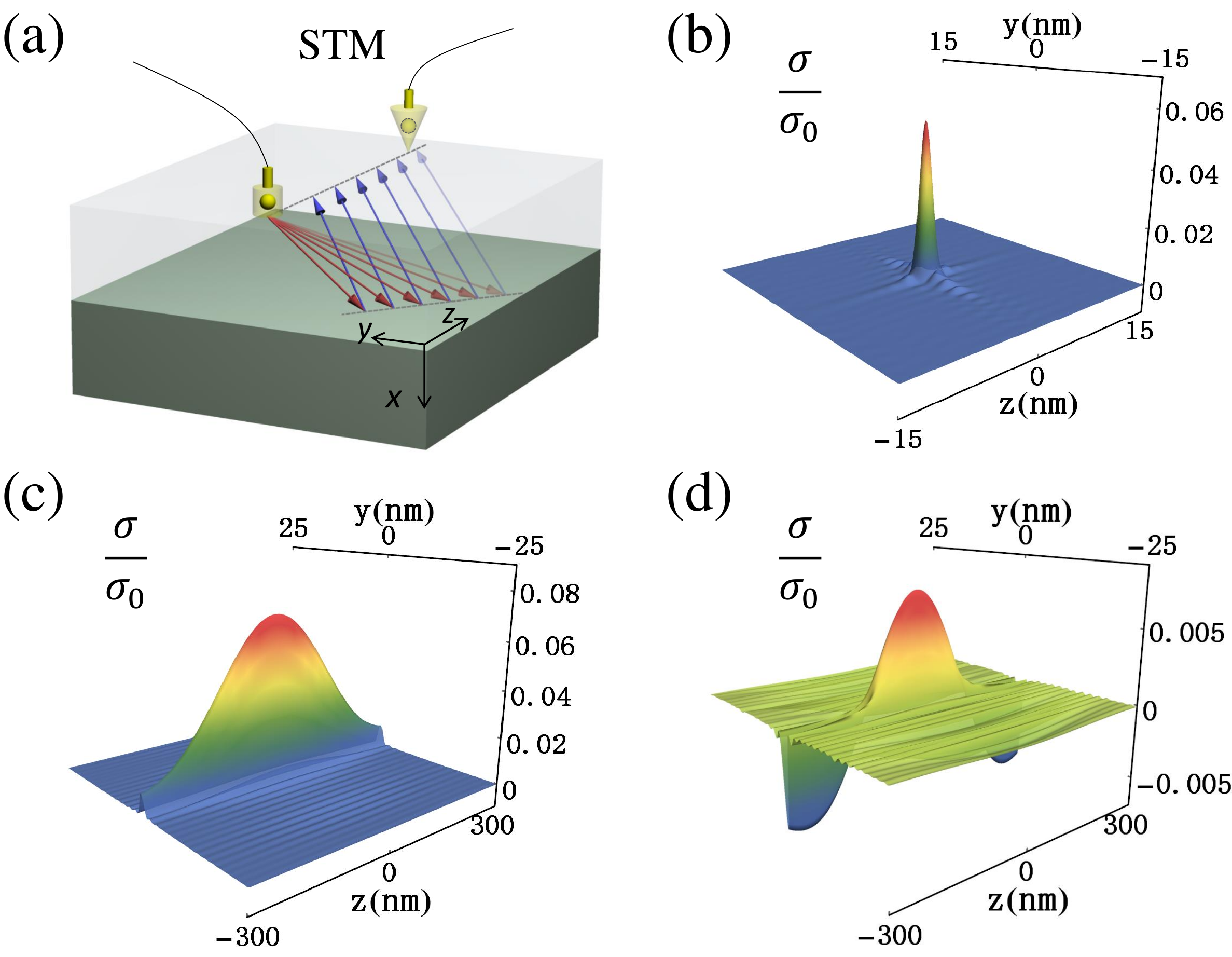}
\caption{(a) Sketch of the two-terminal setup for
nonlocal conductance measurement. (b) Nonlocal conductance in
the RAR regime with $eV=0$ and $\mu=40\Delta$. Nonlocal conductance in the AAR regime for
(c) $Z=0$ and (d) $Z=1$, with $eV=0.2\Delta$ and $\mu=0$.
The location of the open surface of nodal line semimetal is $x_0=200$ nm. All the other parameters
are the same as those in Fig. \ref{Fig3}.}
\label{Fig4}
\end{figure}

Since different types of AR are featured by their
lateral velocities, it is difficult to
identify them by the longitudinal current flow across the
junction. Instead, they can be visibly revealed
by the hole propagation in the $y$-$z$ plane, which manifests
in the nonlocal transport measurement.
Consider the two-terminal setup in Fig. \ref{Fig4}(a)
on top of the nodal line semimetal, which involves
a local electrode at $\bm{x}_l=(-x_{0},0,0)$ and a
movable STM tip at
$\bm{x}_{s}=(-x_{0},y,z)$.
An electron wave packet is injected from the local electrode
which involves contribution of all available ($k_y, k_z$) channels, then gets
Andreev reflected as a hole wave packet, and finally reaches the tip.
For the RAR process, both lateral velocities change sign,
so that the hole retraces the path of the electron, which is localized
in the $y$-$z$ plane. In contrast, only one lateral velocity component is inverted
in the AAR, which indicates that the hole wave packet exhibits a ridge structure,
i.e., localized in one direction while extended in the other [cf. Fig. \ref{Fig4}(a)].

In what follows we calculate the nonlocal conductance using the Green's function
method. The electron coupling between the local terminals and the
nodal line semimetal is described
by  tunneling Hamiltonian
$H_{T}=\sum_{p,\alpha=l,s}t_{\alpha}c_{p\alpha}^{\dagger}\Psi(\bm{x}_{\alpha})+h.c.,$
where $t_{\alpha}$ is the tunneling strength in the $\alpha$ terminal,
$c^{\dagger}_{p\alpha}$ is the creation operator of electron with momentum $p$ in
the $\alpha$ terminal, and
$\Psi(\bm{x})$ is the field operator in the nodal line semimetal.
The nonlocal conductance at zero temperature between the electrode and the STM tip is given by \cite{sm}
\begin{equation}\label{Conductance}
\sigma(eV)=\frac{e^{2}}{h}\sum_{a,b}\zeta_{ab}\mathrm{Tr}[\bm{\Gamma}_{l}^{a}\bm{G}^{R}_{ab}\bm{\Gamma}_{s}^{b}\bm{G}^{A}_{ba}],
\end{equation}
where $\bm{G}^{R,A}$ is the retarded or advanced Green's function,
$\bm{\Gamma}_{l,s}$ is the linewidth function, and subscripts
$a,b=e_{1,2},h_{1,2}$ denote the electron and hole components.
$\zeta_{ab}=-1$ if $a,b$ are both electron/hole states; otherwise, $\zeta_{ab}=1$.
In the weak coupling limit, the correction to the Green's function
due to multiple tunneling can be neglected. Then the retarded Green's function
is constructed by the scattering matrix through \cite{sm}
\begin{equation}\label{GF}
\bm{G}^{R}_{ba}(\bm{x}',\bm{x},E)=-\frac{i}{4\pi^2}\sum_{k_y,k_z}\frac{r_{a}^{b}}
{\hbar v_{a}}e^{i(\gamma_{yz}-\gamma_x)}|b\rangle\langle a|,
\end{equation}
where the phase factors are defined by $\gamma_{yz}=k_y(y'-y)+k_z(z'-z)$ and
$\gamma_x=k_{a} x+k_{b} x'$.
The advanced Green's function
is obtained by $\bm{G}^{A}_{ab}(\bm{x},\bm{x}',E)=[\bm{G}^{R}_{ba}(\bm{x}',\bm{x},E)]^{*}$.
The linewidth function in Eq. \eqref{Conductance} is given by $
\bm{\Gamma}_{\alpha}^{a}(\bm{x}_{1},\bm{x}_{2},E)=2\pi\rho_{\alpha}|t_{\alpha}|^{2}
\delta(\bm{x}_{1}-\bm{x}_{\alpha})\delta(\bm{x}_{2}-\bm{x}_{\alpha})$,
where
$\rho_{\alpha}$ is the density of states at Fermi energy in the $\alpha$ terminal.

We solve Eq. \eqref{Conductance} numerically
and the results of reduced conductance $\sigma/\sigma_0$ are shown in Fig. \ref{Fig4},
with $\sigma_0(eV)=(e^2/h)|2\pi t_lt_s\rho_N|^2\rho_l\rho_s$ and
$\rho_N=2\pi k_0 (\mu+eV)/(\alpha h^2 v_0^2)$ as
the density of states in the nodal line semimetal.
In the intra-band AR regime ($E,\Delta\ll\mu$), the current is dominated by the RAR and the
conductance exhibits a peak structure localized in both $y$ and $z$ directions \cite{note};
see Fig. \ref{Fig4}(b). In the inter-band AR regime
($E,\Delta\gg\mu$), the AAR dominates the nonlocal transport, and
the conductance exhibits a ridge structure
along the $z$ direction as expected; see Fig. \ref{Fig4}(c).
For $U\neq0$, the interface barrier reduces the AAR while enhances
the normal electron reflection [cf. Fig. \ref{Fig3}(f)]. Such an effect
is enhanced with $k_{z}$ increased, and similar anomalous normal reflection $B_{21}$ can even invert the
sign of conductance at the edges of the ridge, as shown in Fig. \ref{Fig4}(d).
Even so, the ridge structural feature still persists, and the effect
of interface barrier can be greatly weakened by setting the energy of incident electron
as close as possible to $\Delta$, where the Andreev resonance
occurs [cf. Fig. \ref{Fig3}(f)].
We conclude that the nonlocal conductance spectra provide
an effective way to identify
different types of AR.

It is worthwhile to discuss the experimental
implementation and potential application of our proposal.
(i) The main building block, Fermi torus has
been reported in doped nodal line semimetals \cite{Emmanouilidou17prb,Takane18npjqm,Hirose20prb,Kwan20prr} as well as the low-density
Rashba gases \cite{Xiang15prb,Landolt12prl,Crepaldi12prl,Cappelluti07prl,Tsutsui12prb}.
(ii) The local electrode
of a few nanometer size can be fabricated by
various state-of-the-art techniques \cite{Kolb97sci,Vieu00ass,Fuechsle12nn}.
The spatially resolved
conductance measurements can be achieved by
a single STM biased with the electrode. The measurements can also be
implemented by the
multi-tip STM technique \cite{Nakayama12am,Li13afm}, which has been
used to perform nonlocal transport measurements \cite{Baringhaus14nat,Just15prl,Hus17prl,Kolmer19nc}.
Both the scale of the local electrode and the accessible spacing
between two STM tips can be much smaller than the
spreading of the ridge structure in the $z$ direction [cf. Fig. \ref{Fig4}(c)]
which ensure a high resolution of the signature.
(iii) High-quality samples of nodal line semimetal
with ultrahigh mobility and clean surface
have been synthesized, which ensures
a long mean free path \cite{Schoop16nc,Sankar17sr,Singha17pnas,Takane18npjqm}. The disorder and
interface roughness can break the
lateral translational symmetry
and lead to diffusion, which will reduce the spatial resolution of the signal.
(iv) Apart from its novel phenomenology,
AAR also provides an effective way to manipulate
Cooper pair splitting process, which may have important applications
in solid-state quantum information processing \cite{Recher01prb,Lesovik01epjb,Hofstetter09nat}.

\begin{acknowledgments}
We thank Fengqi Song, Shaochun Li, Jie Shen, Bin Cheng
and Libo Gao for helpful discussions on the experimental implementation
of our proposal.
This work was supported by the National
Natural Science Foundation of
China under Grant No. 12074172 (W.C.)
and No. 11804130 (W.L.), the startup
grant at Nanjing University (W.C.), the State
Key Program for Basic Researches of China
under Grants No. 2017YFA0303203 (D.Y.X.)
and the Excellent Programme at Nanjing University.
\end{acknowledgments}


\begin{thebibliography}{58}
\bibitem [{not()}]{note}%
  \BibitemOpen
  \href@noop {} {}\bibinfo {note} {In reality, the RAR induced peak structure
  of the conductance is covered by the electrode, so that it cannot be measured
  by the STM tip. Instead, it can be probed directly through the local current
  and revealed in the local conductance spectra.}\BibitemShut {Stop}%
\end{thebibliography}

%

\newpage
\newpage
\onecolumngrid
\renewcommand{\theequation}{S.\arabic{equation}}
\setcounter{equation}{0}
\renewcommand{\thefigure}{S.\arabic{figure}}
\setcounter{figure}{0}

\section{Supplemental Material for ``Anomalous Andreev Reflection on a Torus-Shaped Fermi Surface''}

\author{Wei Luo}
\affiliation{National Laboratory of Solid State Microstructures and School of Physics, Nanjing University, Nanjing 210093, China}
\affiliation{School of Science, Jiangxi University of Science and Technology, Ganzhou 341000, China}

\author{Wei Chen}
\email{Corresponding author: pchenweis@gmail.com}
\affiliation{National Laboratory of Solid State Microstructures and School of Physics, Nanjing University, Nanjing 210093, China}
\affiliation{Collaborative Innovation Center of Advanced Microstructures, Nanjing University, Nanjing 210093, China}

\author{D. Y. Xing}
\affiliation{National Laboratory of Solid State Microstructures and School of Physics, Nanjing University, Nanjing 210093, China}
\affiliation{Collaborative Innovation Center of Advanced Microstructures, Nanjing University, Nanjing 210093, China}


\maketitle

\onecolumngrid
\renewcommand{\theequation}{S.\arabic{equation}}
\setcounter{equation}{0}
\renewcommand{\thefigure}{S.\arabic{figure}}
\setcounter{figure}{0}

\subsection{Transition between different types of Andreev reflection}

\begin{figure}[t!]
\centering
   \includegraphics[width=0.6\columnwidth]{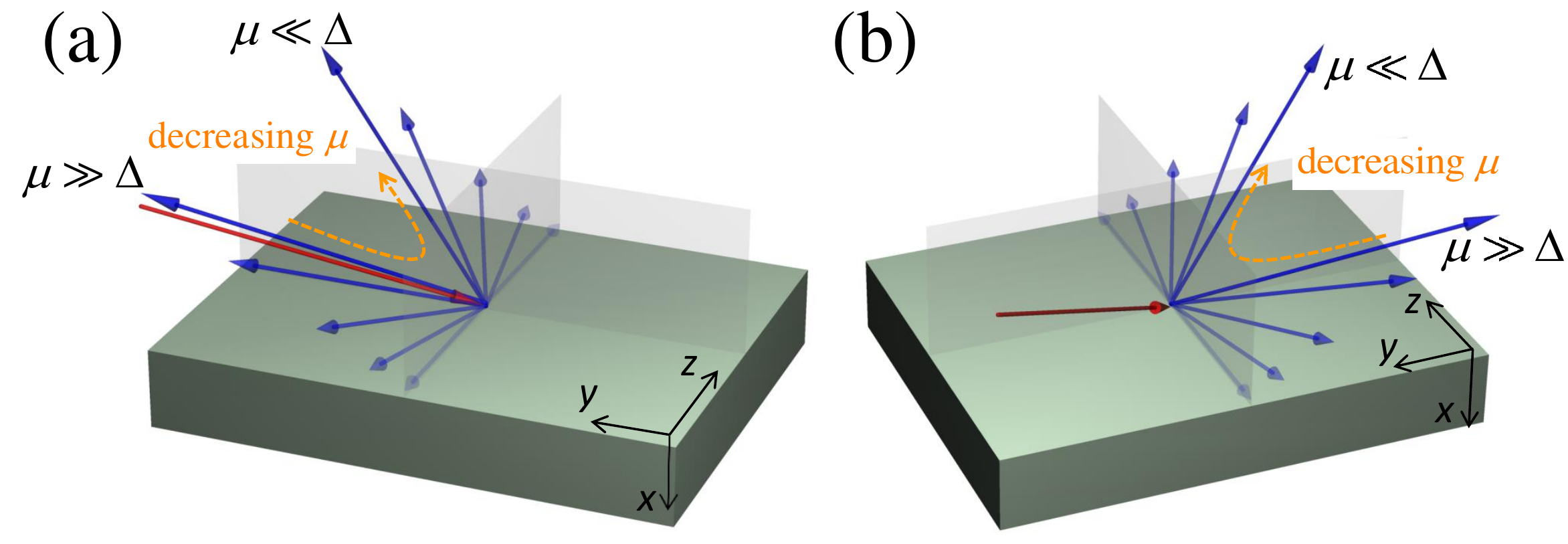}
\caption{(a) RAR-to-AAR transition with $k_x>0$ and (b) AAR-to-SAR transition with $k_x<0$
as the chemical potential $\mu$ decreases.}
\label{figs1}
\end{figure}

Smooth transition of real-space trajectories between different
types of AR is illustrated in Fig. \ref{figs1}. For the dominant AR
processes with $k_x>0$, RAR gradually evolves into AAR
by inverting $v_z$ as the chemical potential
reduces from $\mu\gg\Delta$ to zero. Meanwhile, weak AR processes
with $k_x<0$ undergos a transition from another type of AAR
to SAR by inverting $v_z$ as well.

\subsection{Green's function constructed by the scattering matrix}
Consider a two-terminal system as the junction in the main text, the scattering states
for a particle incident from the left and right side write
\begin{equation}
\begin{split}
|L\rangle_i&=\left\{
              \begin{array}{ll}
                |i\rangle_{\rightarrow}+\sum_j r_{i}^{j}|j\rangle_{\leftarrow}, & \hbox{$x<0$} \\
                \sum_n t_{i}^{n}|n\rangle_{\rightarrow}, & \hbox{$x>0$}
              \end{array}
            \right.\\
|R\rangle_m&=\left\{
                                      \begin{array}{ll}
                                        |m\rangle_\leftarrow+\sum_n {r'}_{m}^{n}|n\rangle_\rightarrow, & \hbox{$x>0$} \\
                                        \sum_{j}{t'}_{m}^{j}|j\rangle_\leftarrow, & \hbox{$x<0$}
                                      \end{array}
                                    \right.
\end{split}
\end{equation}
where $i, n$ $(j, m)$ denote the right-moving (left-moving) modes on the left and right side, respectively.
For a fixed energy, these labels specify the spinor components
such as $|{e}_{1,2}\rangle, |{h}_{1,2}\rangle$ and the momentum of the states.
The Eq. (3) in the main text is a special case of $|L\rangle_i$.

In order to make the scattering matrix unitary, the scattering coefficients should include the ratio of velocity
in the $x$-direction.
We revise the scattering states to
\begin{equation}\label{s}
\begin{split}
|\tilde{L}\rangle_i&=\left\{
              \begin{array}{ll}
                |\tilde{i}\rangle_{\rightarrow}+\sum_j \tilde{r}_{i}^{j}|\tilde{j}\rangle_{\leftarrow}, & \hbox{$x<0$} \\
                \sum_n \tilde{t}_{i}^{n}|\tilde{n}\rangle_{\rightarrow}, & \hbox{$x>0$}
              \end{array}
            \right.\\
|\tilde{R}\rangle_m&=\left\{
\begin{array}{ll}
|\tilde{m}\rangle_\leftarrow+\sum_n \tilde{r}'^{n}_{m}|\tilde{n}\rangle_\rightarrow, & \hbox{$x>0$} \\
\sum_{j}\tilde{t}'^{j}_{m}|\tilde{j}\rangle_\leftarrow, & \hbox{$x<0$}
\end{array}
\right.
\end{split}
\end{equation}
with
\begin{equation}\label{sub}
\begin{split}
|\tilde{i}\rangle&=|i\rangle/\sqrt{\hbar v_i}, \ \ \ \
|\tilde{j}\rangle=|j\rangle/\sqrt{\hbar v_j}, \cdots\\
\tilde{r}_{i}^{j}&=r_{i}^{j}\sqrt{v_j/v_i},
\ \ \tilde{t}_{i}^{n}=t_{i}^{n}\sqrt{v_n/v_i}, \cdots
\end{split}
\end{equation}

In such a way, we obtain the unitary scattering matrix
\begin{equation}
S=\left(
               \begin{array}{cc}
                 \tilde{r} & \tilde{t}' \\
                 \tilde{t} & \tilde{r}' \\
               \end{array}
             \right),\ \ \ \ \
SS^\dag=I.
\end{equation}

We expand the Green's function
under the basis of scattering states \eqref{s}.
For $x,x'<0$, the retarded Green's function is calculated by \cite{Di08}
\begin{equation}\label{g}
\begin{split}
\bm{G}^R(\bm{x}',\bm{x},\omega)=&\langle \bm{x}'|\frac{1}{\omega-H+i0^+}|\bm{x}\rangle\\
=&\sum_i\int dE_{i}\frac{\langle\bm{x}'|\tilde{L}\rangle_i\langle \tilde{L}|_i \bm{x}\rangle}{\omega-E_{i}+i0^+}
+\sum_m\int dE_{m}\frac{\langle \bm{x}'|\tilde{R}\rangle_m\langle \tilde{R}|_m \bm{x}\rangle}{\omega-E_{m}+i0^+}\\
=&\int dE\frac{\sum_i\Big[\tilde{\psi}_i(\bm{x}')+\sum_j \tilde{r}_{i}^{j}\tilde{\psi}_j(\bm{x}')\Big]\Big[\tilde{\psi}^*_i(\bm{x})+\sum_{j'} \tilde{r}_{i}^{j'*}\tilde{\psi}^*_{j'}(\bm{x})\Big]+\sum_m\Big[\sum_j \tilde{t}'^{j}_{m}\tilde{\psi}_j(\bm{x}')\Big]\Big[\sum_{j'}t'^{j'*}_{m}\tilde{\psi}^*_{j'}(\bm{x})\Big]}{\omega-E+i0^+}\\
\bm{G}^R(\bm{x}',\bm{x},\omega)=&\int dE\frac{\sum_i\tilde{\psi}_i(\bm{x}')\tilde{\psi}^*_i(\bm{x})+\sum_{j}\tilde{\psi}_j(\bm{x}')
\tilde{\psi}^*_{j}(\bm{x})}{\omega-E+i0^+}+\int dE\frac{\sum_{ij}\tilde{r}^{j*}_{i}\tilde{\psi}_i(\bm{x}')
\tilde{\psi}^*_{j}(\bm{x})+\tilde{r}_{i}^{j}\tilde{\psi}_j(\bm{x}')\tilde{\psi}^*_i(\bm{x})}{\omega-E+i0^+}.\\
\end{split}
\end{equation}
In the last step, we have used the unitary condition $\sum_{jj'}(\tilde{t}'\tilde{t}'^{\dagger}+\tilde{r}\tilde{r}^{\dagger})=\delta_{jj'}$.
The first term of Eq. \eqref{g} describes the free propagation of particle,
which has no contribution to the nonlocal transport between
the local electrode and the STM tip and thus can be dropped. Calculating
the remaining term using the residue theorem yields
\begin{equation}
\begin{split}
\bm{G}^R(\bm{x}',\bm{x},\omega)=&-2\pi i\sum_{ij} \tilde{r}_{i}^{j}\tilde{\psi}_j(\bm{x}')\tilde{\psi}^*_i(\bm{x})
\end{split}
\end{equation}
The wave functions take the form of
\begin{equation}
\begin{split}
\tilde{\psi}_i(\bm{x})&=\frac{1}{\sqrt{2\pi\hbar v_i}}\chi_{i}(y,z)e^{ik_ix},\\
\tilde{\psi}_j(\bm{x}')&=\frac{1}{\sqrt{2\pi\hbar v_j}}\chi_{j}(y',z')e^{-ik_jx'},
\end{split}
\end{equation}
with the transverse modes being labeled
by the momentum $k_y, k_z$ as
\begin{equation}
\begin{split}
\chi_{i}(y,z)&=\frac{1}{2\pi} |a\rangle e^{i(k_yy+k_zz)},\\
\chi_{j}(y',z')&=\frac{1}{2\pi}|b\rangle e^{i(k_yy'+k_zz')},
\end{split}
\end{equation}
where $|a\rangle, |b\rangle=|{e}_{1,2}\rangle,|{h}_{1,2}\rangle$ are the spinor part of the wave
function.

Inserting the wave functions into the expression of Green's function and given the conservation of transverse momentum
during scattering yields
\begin{equation}\label{gf}
\begin{split}
\bm{G}^{R}_{ba}(\bm{x}',\bm{x},E)&=-\frac{i}{4\pi^2}\sum_{k_y,k_z}\frac{r_{a}^{b}}
{\hbar v_{a}}e^{i\big[k_y(y'-y)+k_z(z'-z)\big]}e^{-i\big[k_{a} x+k_{b} x'\big]}|b\rangle\langle a|,
\end{split}
\end{equation}
which is Eq. (7) in the main text, with $k_{a,b}=k_{a,b}(k_y,k_z,E), v_{a,b}=v_{a,b}(k_y,k_z,E)$

\subsection{Green's function formula of nonlocal conductance}
Consider the local electrode and STM tip coupled with the nodal line semimetal as shown
in Fig. 4(a) of the main text. The whole system is composed of two local terminals and the
central scattering region: the nodal line semimetal-superconductor junction.
Following Ref. \cite{Jauho94prb}, we derive the general expression
of the nonlocal conductance for the system which contains a superconductor.
The whole Hamiltonian in Nambu space is
\begin{equation}
\begin{split}
H_{L}&=\sum_{\alpha=l,s}\sum_{p}\tilde{c}^\dag_{p\alpha}h_{p\alpha}\tilde{c}_{p\alpha},\\
H_{0}&=\sum_{mn}\tilde{d}^\dag_m h^s_{mn}\tilde{d}_n,\\
H_{T}&=\sum_{\alpha=l,s}\sum_{p,n}[\tilde{c}^\dag_{p\alpha}t_{p\alpha,n}\tilde{d}_n+H.c.],\\
\end{split}
\end{equation}
where $H_L$ is for the local electrode and STM tip, $H_0$ corresponds to
the nodal line semimetal-superconductor junction and $H_T$ is the coupling between them.
The single-particle Hamiltonian is defined as $h_{p\alpha}=\left(
       \begin{array}{cc}
         \epsilon_{p\alpha} & 0 \\
          0 & -\epsilon_{-p\alpha} \\
       \end{array}
      \right)=\left(
        \begin{array}{cc}
         \epsilon^e_{p\alpha} & 0 \\
          0 & \epsilon^h_{p\alpha} \\
        \end{array}
     \right)$,
$t_{p\alpha,n}=\left(
       \begin{array}{cc}
         T_{p\alpha,n} & 0 \\
          0 & -T^{*}_{-p\alpha,n} \\
       \end{array}
      \right)=\left(
        \begin{array}{cc}
         t^{e}_{p\alpha,n} & 0 \\
          0 & t^{h}_{p\alpha,n} \\
        \end{array}
     \right)$,
and the Nambu spinors are defined by $\tilde{c}_{p\alpha}=(\tilde{c}^{e}_{p\alpha},\tilde{c}^{h}_{p\alpha})^{T}
=(c_{p\alpha\uparrow},c^{\dagger}_{-p\alpha\downarrow})^{T}$,
$\tilde{d}_{n}=(\tilde{d}^{e}_{n},\tilde{d}^{h}_{n})^{T}
=(d_{n\uparrow},d^{\dagger}_{n\downarrow})^{T}$. All energies and coupling stength are
time-independent.

The current in the STM tip is defined by
\begin{equation}
\begin{split}
J_{s}(t)&=-e\langle\dot{N}_{s}\rangle=-\frac{ie}{\hbar}\langle[H,N_{s}]\rangle\\
&=\frac{ie}{\hbar}\sum_{pn}[\langle \tilde{c}^\dag_{ps}(t)\tau_zt_{ps,n} \tilde{d}_{n}(t)\rangle-\langle \tilde{d}_{n}^\dag (t)\tau_zt^*_{ps,n}\tilde{c}_{ps}(t)\rangle].
\end{split}
\end{equation}
Define the lesser Green's fuction
\begin{eqnarray}
G^{<\mu\nu}_{n,p\alpha}(t,t')\equiv i\langle\tilde{c}^{\dag \nu}_{p\alpha}(t')\tilde{d}^\mu_n(t)\rangle,
\end{eqnarray}
where the superscripts  $\mu,\nu$ denote the electron-hole components.
Using the Green's function, the current is expressed as
\begin{equation}\label{current}
J_{s}(t)=\frac{2e}{\hbar}\text{Re}\Big\{\sum_{pn}\text{Tr}[\tau_zt_{ps,n}(t)\bm{G}^<_{n,ps}(t,t)]\Big\}.
\end{equation}
The lesser Green's function can be obtained
by analytic continuation on the following contour-ordered Green's function
\begin{equation}\label{cgf}
\begin{split}
G^{\mu\nu}_{n,p\alpha}(\tau,\tau')&=\sum_{m}\int\tau_1G^{\mu\nu}_{nm}(\tau,\tau_1)
t^{\nu*}_{p\alpha,m}(\tau_1)g^{\nu}_{p\alpha}(\tau_1,\tau'),
\end{split}
\end{equation}
which yields \cite{Jauho94prb}
\begin{equation}\label{cgf2}
G^{<\mu\nu}_{n,p\alpha}(t,t')=\sum_{m}\int dt_1 \Big[G^{R\mu\nu}_{nm}(t,t_1)t^{\nu*}_{p\alpha,m}(t_1)g^{<\nu}_{p\alpha}(t_1,t')
+G^{<\mu\nu}_{nm}(t,t_1)t^{\nu*}_{p\alpha,m}(t_1)g^{A\nu}_{p\alpha}(t_1,t')\Big].
\end{equation}
The bare Green's function in the lead is diagonal in Nambu space, which is
\begin{equation}\label{bare green}
\begin{split}
g^{<\nu}_{p\alpha}(t,t')&=i\langle \tilde{c}^{\dag \nu}_{p\alpha}(t')\tilde{c}^\nu_{p\alpha}(t)\rangle
=if_\alpha^\nu(\epsilon^{\nu}_{p\alpha})e^{-i\epsilon^\nu_{p\alpha}(t-t')},\\
g^{R,A\nu}_{p\alpha}(t,t')&=\mp i\theta(\pm t\mp t')\langle[\tilde{c}^\nu_{p\alpha}(t),\tilde{c}^{\dag \nu}_{p\alpha}(t')]_+\rangle
=\mp i\theta(\pm t\mp t')e^{-i\epsilon^\nu_{p\alpha}(t-t')}.
\end{split}
\end{equation}
For the $\alpha$ terminal, the hole distribution function $f_\alpha^h$ is related to that of electron
via $f^h_\alpha(\epsilon^{h}_{p\alpha})=1-f^e_\alpha(\epsilon_{-p\alpha})$.
As the electrode is biased by a voltage $eV$, the electron distribution is
$f^e_\alpha(\omega)=f_0(\omega-eV)$ and the hole distribution is
$f^h_\alpha(\omega)=f_0(\omega+eV)$, with $f_0(\omega)=1/(e^{\beta\omega}+1)$
being the Fermi-Dirac distribution function.
The current Eq. \eqref{current} includes only the diagonal elements [$\nu=1,2$ $(e,h)$] of the Green's function,
which can be labeled by a single superscript $\nu$ as
\begin{equation}
\begin{split}
G^{<\nu}_{n,p\alpha}(t,t')=\sum_{m}\int dt_1 t^{\nu*}_{p\alpha,m}\Big[G^{R\nu}_{nm}(t,t_1)if_\alpha^\nu(\epsilon^{\nu}_{p\alpha})e^{-i\epsilon^{\nu}_{p\alpha}(t_{1}-t')}
+G^{<\nu}_{nm}(t,t_1)i\theta(-t_1+t')e^{-i\epsilon^\nu_{p\alpha}(t_{1}-t')}\Big].
\end{split}
\end{equation}
Then the current reduces to
\begin{equation}
\begin{split}
J_{s}(t)
&=\frac{2e}{\hbar}\text{Im}\Big\{\sum_{pnm\nu}(-1)^{\nu}
t^\nu_{ps,n}\int_{-\infty}^t dt_1 t^{\nu*}_{ps,m}e^{-i\epsilon^\nu_{ps}(t_{1}-t)}
\Big[G^{R\nu}_{nm}(t,t_1)f^\nu_{s}(\epsilon^{\nu}_{ps})+G^{<\nu}_{nm}(t,t_1)\Big]
\Big\}\\
&=\frac{2e}{\hbar}\text{Im}\Big\{\sum_{nm\nu}(-1)^\nu\int d\epsilon_{s}^{\nu}
\rho^{\nu}_{s}(\epsilon_{s}^{\nu})t^\nu_{s,n}(\epsilon_{s}^{\nu})
\int_{-\infty}^t dt_1 e^{-i\epsilon^\nu_{s}(t_{1}-t)} t^{\nu*}_{s,m}(\epsilon_{s}^{\nu})
\Big[G^{R\nu}_{nm}(t-t_1)f_s^\nu(\epsilon^{\nu}_{s})
+G^{<\nu}_{nm}(t-t_1)\Big]\Big\}\\
&=\frac{2e}{\hbar}\text{Im}\Big\{\sum_{nm\nu}(-1)^\nu\int_{-\infty}^{0} dt_1\int \frac{d\epsilon}{2\pi} e^{-i\epsilon t_1}
[\Gamma^{\nu}_{s}(\epsilon)]_{mn} \Big[G^{R\nu}_{nm}(-t_1)f^\nu_{s}(\epsilon)
+G^{<\nu}_{nm}(-t_1)\Big]\Big\},
\end{split}
\end{equation}
where the linewidth function $[\Gamma^{\nu}_{\alpha}(\epsilon)]_{mn}=2\pi\rho^\nu_{\alpha}(\epsilon)
t^{\nu*}_{\alpha,m}(\epsilon)t^\nu_{\alpha,n}(\epsilon)$ in the $\alpha$ terminal satisfies
$[\Gamma^{\nu}_{\alpha}(\epsilon)]_{nm}^{*}=[\Gamma^{\nu}_{\alpha}(\epsilon)]_{mn}$,
and $\rho_\alpha^\nu$ is the density of states.
The current is expressed by the Green's function in the region of the junction.
It can be written in a more compact form as
\begin{equation}
\begin{split}
J_{s}(t)&=\frac{2e}{\hbar}\sum_{\nu}(-1)^\nu\int_{-\infty}^{0} dt_1\int
\frac{d\epsilon}{2\pi} \text{Im}\Big\{ \text{Tr}\Big\{e^{-i\epsilon t_1}
\bm{\Gamma}_{s}^{\nu}(\epsilon) \Big[\bm{G}^{R}_\nu(-t_1)f_{s}^\nu(\epsilon)
+\bm{G}^{<}_\nu(-t_1)\Big]\Big\}\Big\},
\end{split}
\end{equation}
where we have changed the superscript $\nu$ of the Green's function to the subscript.
Using $\text{Tr}[\bm{\Gamma}\bm{G}]^*=\text{Tr}[\bm{\Gamma}^\dag\bm{G}^\dag]$ we have
\begin{equation}\label{generalnambucurrent}
\begin{split}
J_{s}(t)
=&\frac{e}{i\hbar}\sum_{\nu}(-1)^\nu\int \frac{d\epsilon}{2\pi}\int_{-\infty}^0 dt_1 \Big\{e^{-i\epsilon t_1}f^\nu_{s}(\epsilon)
\text{Tr}\Big[\bm{\Gamma}_{s}^{\nu}(\epsilon)\bm{G}^{R}_\nu(-t_1)\Big]-e^{i\epsilon t_1}f^\nu_{s}(\epsilon)
\text{Tr}\Big[\Big(\bm{\Gamma}_{s}^{\nu}(\epsilon)\Big)^\dag\Big(\bm{G}^{R}_\nu(-t_1)\Big)^\dag\Big]\\
&+e^{-i\epsilon t_1}
\text{Tr}\Big[\bm{\Gamma}_{s}^{\nu}(\epsilon)\bm{G}^{<}_{\nu}(-t_1)\Big]-e^{i\epsilon t_1}
\text{Tr}\Big[\Big(\bm{\Gamma}_{s}^{\nu}(\epsilon)\Big)^\dag\Big(\bm{G}^{<}_{\nu}(-t_1)\Big)^\dag\Big]\Big\}\\
=&\frac{e}{i\hbar}\sum_{\nu}(-1)^\nu\int \frac{d\epsilon}{2\pi} \text{Tr}\Big\{\bm{\Gamma}_{s}^{\nu}(\epsilon)\Big(\bm{G}^{<}_{\nu}(\epsilon)+
f_{s}^\nu(\epsilon)\Big[\bm{G}^{R}_{\nu}(\epsilon)-\bm{G}^{A}_{\nu}(\epsilon)\Big]\Big)\Big\}.
\end{split}
\end{equation}
The lesser Green's function can be solved by the Keldysh equation
\begin{equation}
\begin{split}
\bm{G}^{<}_{\nu}(\epsilon)&=\sum_\mu\bm{G}^{R}_{\nu\mu}(\epsilon)\bm{\Sigma}^{<}_{\mu\mu}(\epsilon)\bm{G}^{A}_{\mu\nu}(\epsilon),\\
\end{split}
\end{equation}
where the self-energy
\begin{equation}
\Sigma^{<\mu\mu}_{nm}(\epsilon)=\sum_{p\alpha=l,s}t_{p\alpha,n}^{\mu*}
g^{<\mu}_{p\alpha}(\epsilon)t^{\mu}_{p\alpha,m}
\end{equation}
is due to the coupling with the electrode
and the STM tip, which is diagonal in the Nambu space.
Using Eq.\ (\ref{bare green}) we have
\begin{equation}
\begin{split}
g^{<\nu}_{p\alpha}(\epsilon)&=\int dt e^{i\epsilon t} g^{<\nu}_{p\alpha}(t)=2\pi i f^\nu_\alpha(\epsilon^\nu_{p\alpha})\delta(\epsilon-\epsilon^\nu_{p\alpha}),\\
g^{R,A\nu}_{p\alpha}(\epsilon)&=\int dt e^{i\epsilon t} g^{R,A\nu}_{p\alpha}(t)=
\frac{1}{\epsilon-\epsilon^\nu_{p\alpha}\pm i0^+},
\end{split}
\end{equation}
which yields
\begin{equation}
\begin{split}
\Sigma^{<\nu\nu}_{nm}(\epsilon)
&=i [(\Gamma^{\nu}_{s})_{nm}f^\nu_{s}(\epsilon)+
(\Gamma^{\nu}_{l})_{nm}f^\nu_{l}(\epsilon)],\\
\Sigma^{R,A\nu\nu}_{nm}(\epsilon)
&=\mp\frac{i}{2}[(\Gamma_{s}^{\nu})_{nm}+(\Gamma_{l}^{\nu})_{nm}],
\end{split}
\end{equation}
or formally,
\begin{equation}
\begin{split}
\bm{\Sigma}^{<}_{\nu}(\epsilon)&=i [\bm{\Gamma}_{s}^{\nu}(\epsilon)f^\nu_{s}(\epsilon)+\bm{\Gamma}_{l}^{\nu}(\epsilon)f^\nu_{l}(\epsilon)],\\
\bm{\Sigma}^{R,A}_{\nu}(\epsilon)&=\mp\frac{i}{2}[\bm{\Gamma}_{s}^{\nu}(\epsilon)+\bm{\Gamma}_{l}^{\nu}(\epsilon)].
\end{split}
\end{equation}
Applying the Dyson equation
\begin{equation}
[\bm{G}^{R,A}]^{-1}=[\bm{g}^{R,A}]^{-1}-\bm{\Sigma}^{R,A},
\end{equation}
we have
\begin{equation}
[\bm{G}^{A}]^{-1}-[\bm{G}^{R}]^{-1}=\bm{\Sigma}^R-\bm{\Sigma}^A,
\end{equation}
and then
\begin{equation}
\begin{split}
\bm{G}^R(\epsilon)-\bm{G}^A(\epsilon)&=\bm{G}^R(\epsilon)[\bm{\Sigma}^R(\epsilon)-\bm{\Sigma}^A(\epsilon)]\bm{G}^A(\epsilon).\\
\end{split}
\end{equation}
The diagonal elements are
\begin{equation}
\bm{G}^{R}_{\nu}(\epsilon)-\bm{G}^{A}_{\nu}(\epsilon)
=\sum_\mu-i\bm{G}^{R}_{\nu\mu}(\epsilon)[\bm{\Gamma}_{l}^{\mu}(\epsilon)
+\bm{\Gamma}_{s}^{\mu}(\epsilon)]\bm{G}^{A}_{\mu\nu}(\epsilon).
\end{equation}
Inserting the lesser self-energy $\bm{\Sigma}^<$ and the relation
above into Eq. \eqref{generalnambucurrent} yields
\begin{equation}\label{lb}
\begin{split}
J_{s}
&=\frac{e}{\hbar}\sum_{\nu\mu}(-1)^\nu\int\frac{d\epsilon}{2\pi}
\text{Tr}\Big\{\bm{\Gamma}_{s}^{\nu}(\epsilon)\Big(\bm{G}^{R}_{\nu\mu}(\epsilon)\Big[
\bm{\Gamma}_{s}^{\mu}(\epsilon)f_{s}^\mu(\epsilon)+\bm{\Gamma}_{l}^{\mu}(\epsilon)
f^\mu_{l}(\epsilon)\Big]\bm{G}^{A}_{\mu\nu}(\epsilon)
-f_{s}^{\nu}(\epsilon)\bm{G}^{R}_{\nu\mu}(\epsilon)\Big[\bm{\Gamma}_{s}^{\mu}(\epsilon)
+\bm{\Gamma}_{l}^{\mu}(\epsilon)\Big]
\bm{G}^{A}_{\mu\nu}(\epsilon)\Big)\Big\}\\
&=\frac{e}{h}\sum_\nu (-1)^\nu\int d\epsilon \text{Tr}\Big\{
(f_{s}^{\bar{\nu}}-f_{s}^\nu)\bm{\Gamma}_{s}^{\nu}\bm{G}^{R}_{\nu\bar{\nu}}
\bm{\Gamma}_{s}^{\bar{\nu}}\bm{G}^{A}_{\bar{\nu}\nu}
+(f^{\nu}_{l}-f_{s}^\nu)\bm{\Gamma}_{s}^{\nu}\bm{G}^{R}_{\nu\nu}
\bm{\Gamma}_{l}^{\nu}\bm{G}^{A}_{\nu\nu}
+(f^{\bar{\nu}}_{l}-f^\nu_{s})\bm{\Gamma}_{s}^{\nu}\bm{G}^{R}_{\nu\bar{\nu}}
\bm{\Gamma}_{l}^{\bar{\nu}}\bm{G}^{A}_{\bar{\nu}\nu}
\Big\}.
\end{split}
\end{equation}
Assume that the local electrode is biased with the STM tip
and the superconductor with energy $eV$, the zero-temperature differential conductance
$\sigma(eV)=\partial J_s/\partial V$ is given by
\begin{equation}\label{cond}
\begin{split}
\sigma&=
\frac{e^{2}}{h}\sum_{a,b}\zeta_{ab}\mathrm{Tr}[\bm{\Gamma}_{s}^{a}\bm{G}^{R}_{ab}\bm{\Gamma}_{l}^{b}\bm{G}^{A}_{ba}],
\end{split}
\end{equation}
where $\zeta_{ab}=-1$ if $a,b$ are both electron/hole states; otherwise,
$\zeta_{ab}=1$.

In the present system,
the tunneling Hamiltonian is
\begin{equation}
H_{T}=\sum_{p,\alpha=l,s}t_{\alpha}c_{p\alpha}^{\dagger}\Psi(\bm{x}_{\alpha})+h.c.,
\end{equation}
and the corresponding linewidth function is
\begin{equation}
\bm{\Gamma}_{\alpha}^{a}(\bm{x}_{1},\bm{x_{2}},E)=2\pi\rho_{\alpha}^{a}|t_{\alpha}|^{2}
\delta(\bm{x}_{1}-\bm{x}_{\alpha})\delta(\bm{x}_{2}-\bm{x}_{\alpha}),
\end{equation}
with $\bm{x}_{l(s)}$ the location of local electrode (STM tip). Under
the tunneling limit in our setup, the multiple tunneling has
negligible contribution to the nonlocal current.
Then we can insert Eq. \eqref{gf} into the conductance formula \eqref{cond} and obtain
\begin{equation}
\begin{split}
\sigma(eV)&=\frac{e^{2}}{h}\sum_{a,b}\zeta_{ab}\int d\bm{x}_{1}\bm{x}_{2}\bm{x}_{3}\bm{x}_{4}
\bm{\Gamma}_{l}^{a}(\bm{x}_{1},\bm{x}_{2},E)\bm{G}^{R}_{ab}(\bm{x}_{2},\bm{x}_{3},E)
\bm{\Gamma}_{s}^{b}(\bm{x}_{3},\bm{x}_{4},E)\bm{G}^{A}_{ba}(\bm{x}_{4},\bm{x}_{1},E)\\
&=\frac{e^{2}}{h}\sum_{a,b}\zeta_{ab}|2\pi t_lt_s|^2\rho_{l}\rho_{s}
\bm{G}^{R}_{ab}(\bm{x}_{l},\bm{x}_{s},E)
\bm{G}^{A}_{ba}(\bm{x}_{s},\bm{x}_{l},E)\\
&=\sigma_{0}(eV)\sum_{a,b}\zeta_{ab}\Big|\int dk_{y}dk_{z}\frac{1}{4\pi^2}\frac{\rho_{k_{y},k_{z}}}{\rho_{N}}r_{ba}^{0}e^{i[k_y(y'-y)+k_z(z'-z)]}e^{-i[k_{a} x+k_{b} x']}\Big|^{2}.
\end{split}
\end{equation}
with $\sigma_0(eV)=(e^2/h)|2\pi t_lt_s\rho_N|^2\rho_l\rho_s$ and $\rho_N$ is the density of states in nodal line semimetal.

\subsection{Tunneling conductance of the junction}

\begin{figure}[t!]
 \centering
   \includegraphics[width=0.6\columnwidth]{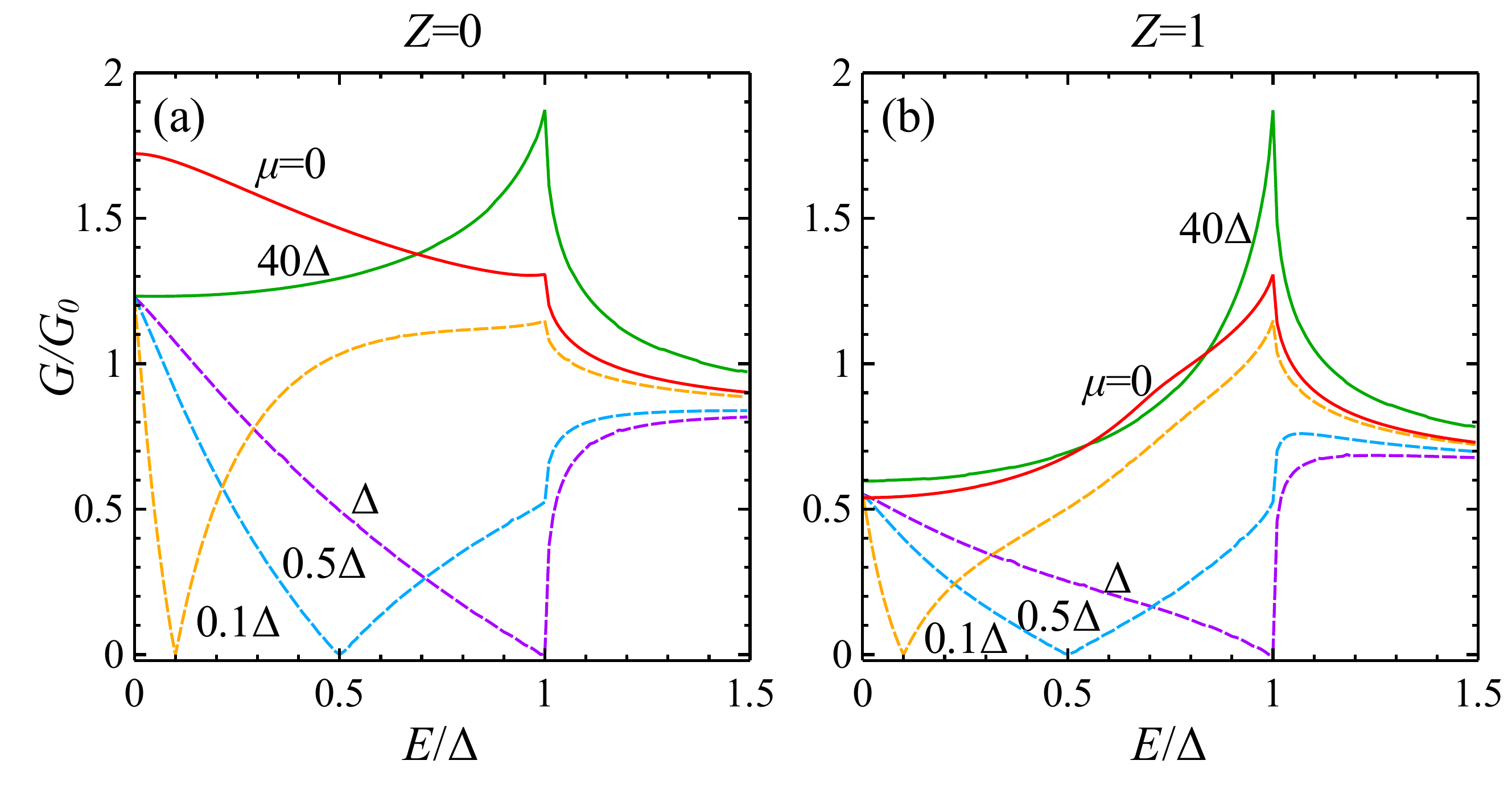}
\caption{Conductance spectra for (a) zero and (b) finite interface
barrier with different chemical potentials. All relevant parameters
are the same as those in Fig. 3 in the main text.}
\label{figs2}
\end{figure}

In this section, we calculate the longitudinal conductance
across the junction. Inserting the reflection coefficients
into the Blonder-Tinkham-Klapwijk formula \cite{Blonder82prb} yields
\begin{equation}
G(eV)=\frac{e^2 S}{4\pi^2h}\int dk_y dk_z\Big[2+\sum_{i,j=1,2}(A_{ji}-B_{ji})\Big],
\end{equation}
where $S$ is the cross-section area of the junction.
The numerical results of the reduced conductance $G/G_0$
are shown in Fig. \ref{figs2},
where $G_0(eV)=G(eV)|_{A_{ji}=B_{ji}=0}$ is the
conductance of the uniform nanowire along the $x$ direction.
For zero interface barrier in Fig. \ref{figs2}(a),
the conductance spectra resemble those of graphene \cite{Beenakker06prl}
but with a different physical meaning. Here, an AAR-to-RAR crossover takes
place as the chemical potential increases, which is different from the SAR-to-RAR
transition in graphene. Moreover, for a finite interface barrier
in Fig. \ref{figs2}(b),
even the conductance spectra of
AAR and RAR become similar and cannot be discriminated.
We conclude that different types of AR cannot be identified by the tunneling
conductance spectra.

\end{document}